# Comparison of crystal structures and effects of Co substitution in a new member of Fe-1111 superconductor family *Ae*FeAsF (*Ae* = Ca and Sr): a possible candidate for higher-$T_c$ superconductor

Takatoshi Nomura[1], Yasunori Inoue[1], Satoru Matsuishi[2,] Masahiro Hirano[2,3], Jung Eun Kim[4], Kenichi Kato[5], Masaki Takata[4,5] and Hideo Hosono[1,2,3]

[1] Materials and Structure Laboratory, Tokyo Institute of Technology, 4259 Nagatsuta-cho, Midori-ku, Yokohama 226-8503, Japan
[2] Frontier Research Center, Tokyo Institute of Technology, 4259 Nagatsuta-cho, Midori-ku Yokohama 226-8503, Japan
[3] ERATO-SORST, JST, Frontier Research Center, Tokyo Institute of Technology, 4259 Nagatsuta-cho, Midori-ku ,Yokohama 226-8503, Japan
[4] Japan Synchrotron Radiation Research Institute, 1-1-1, Kouto, Sayo-cho, Sayo-gun, Hyogo 679-5198, Japan
[5] RIKEN SPring-8 Center, 1-1-1, Kouto, Sayo-cho, Sayo-gun, Hyogo 679-5148, Japan

Corresponding E-mail: hosono@msl.titech.ac.jp

**Abstract.**

We refined crystal structures of newly found members of the Fe-1111 superconductor family, CaFe$_{1-x}$Co$_x$AsF and SrFe$_{1-x}$Co$_x$AsF ($x$ = 0, 0.06, 0.12) by powder synchrotron X-ray diffraction analysis. The tetragonal to orthorhombic phase transitions were observed at ~120 K for unsubstituted CaFeAsF and at ~180 K for unsubstituted SrFeAsF, the transition temperatures agreeing with kinks observed in temperature-dependent resistivity curves. Although the transition temperature decreases, the structural phase transitions were observed below 100 K in both samples of $x$ = 0.06, and finally they were suppressed in the doping level of $x$ = 0.12. The refined structures reveal that distortions of the FeAs$_4$ tetrahedron from the regular tetrahedron likely originate from mismatches in atomic radii among the constituent elements. In this system, the enlarged FeAs$_4$ tetrahedron resulting from larger radius of Sr than that of Ca is flattened along *a-b* plane, whereas the smaller radius of Ca makes the tetrahedron closer to regular one, and their characteristic shapes are further enhanced by Co substitution. These results suggest that the CaFeAsF compound is a promising candidate for higher-$T_c$ superconductor.

## 1. Introduction

Since the first report of superconductivity in F-substituted LaFeAsO with $T_c$ = 26 K much progress have been made in iron-based superconductors both experimentally and theoretically. However, many important material and physical issues still remain controversial. The progress includes enhancement of $T_c$ up to ~55 K [2–4], versification in the superconductor family members, development of carrier doping technique, and fundamental understanding of underlining physics. These efforts have clarified that the parent compounds for the superconductors involve a two-dimensional square lattice of Fe element tetrahedrally coordinated by four As or chalcogenide ions (Se and Te) as a key building block. It is most likely the parent compounds are "bad metal" due to the association of multiple Fe $3d$ orbital branches with the Fermi surface to form multiple electron and hole pockets on it. It is also acknowledged that the high-$T_c$ superconductors are realized by doping of electron or hole carriers in the Fe lattice either by direct or indirect doping technique. The former technique is achieved, for instance, by a partial replacement of Fe with Co in $BaFe_2As_2$ (Fe-122 compound) [5] or formation of vacancy in the Se site in PbO-type FeSe (Fe-11 compound) [6]. On the other hand, the latter technique is realized by substitution of $F^-$ ion to the $O^{2-}$ site in the blocking LaO layer in LaFeAsO (Fe-1111 compound) or formation of vacancy in the Li blocking layer in LiFeAs (Fe-111 compound) [7].

A prominent feature in these superconductors is that the parent compounds undergo a structural phase transition at around 100–200 K, which is accompanied by an antiferromagnetic order of Fe spins. These transitions can be observed as an "anomaly" in temperature-dependent electrical resistivity ($\rho$) and a small decrease in magnetic susceptibility ($\chi$). They are further confirmed by microscopic measurements including Neutron scattering (ND) [8], X-ray diffraction (XRD) [9], Nuclear Magnetic Resonance (NMR) [10] and Mossbauer spectroscopy [11], demonstrating a good agreement of the starting temperature of the symmetry lowering transition with as a kink in the $\rho$-$T$ curve ($T_{anom}$). It shifts to the lower temperature side by the carrier doping and disappears with a further increase in the doping level, leading to eventual appearance of the superconductivity. Such behaviors have not been observed in low-$T_c$ superconductors with similar crystal structures (< ~4 K) such as $Ln$FePO and $Ln$NiAsO ($Ln$ = rare-earth elements) [12–14], indicating a close associate of the transitions with the high $T_c$.

Matsuishi et al. have recently demonstrated a new family member of Fe-1111 compound, $Ae$FeAsF ($Ae$ = Ca or Sr), with $Ae$F in place of $Ln$O as the blocking layer, which exhibits superconductivity as a result of the partial replacement of Fe with Co [15,16]. $CaFe_{1-x}Co_xAsF$ undergo superconducting transition for a wide Co concentration ($x$) from 0.06 to 0.30 with the maximum $T_c$ of ~22 K at $x$ ~0.10, whereas $SrFe_{1-x}Co_xAsF$ becomes superconductor only in a narrow $x$ range and relatively low $T_c$ (~7 K) has been observed for $x$ = 0.125. The structural and magnetic phase transitions have been also confirmed in unsubstituted SrFeAsF, whose feature is quite similar to those of $Ln$FeAsO [17].

In this study, we examine synchrotron XRD measurements on $CaFe_{1-x}Co_xAsF$ and $SrFe_{1-x}Co_xAsF$ ($x$ = 0, 0.06, 0.12) from 30 K to 300 K, and demonstrate that the structural phase transitions take place at

~120 K for unsubstituted CaFeAsF and ~180 K for unsubstituted SrFeAsF. It is noteworthy that CaFe$_{1-x}$Co$_x$AsF, whose $T_{\text{anom}}$ is lower than that of SrFe$_{1-x}$Co$_x$AsF exhibits much higher $T_c$. Further, the crystal structures of CaFeAsF and SrFeAsF are refined by Rietveld analyses on the XRD data and differences in them are discussed in relation to their $T_c$ values. Effects of the Co substitution on the structures are also evaluated including chemical bond lengths, layer distance and magnitude of distortions of the FeAs$_4$ and $Ae_4$F tetrahedrons from the regular tetrahedron.

## 2. Experimental & Analysis methods

Unsubstituted $Ae$FeAsF samples were prepared by a solid state reaction of an 1 : 1 : 1 mixture of Fe$_2$As, Ca(Sr)As and Ca(Sr)F$_2$ powders. For Co-substituted samples, 6% and 12% of the Fe$_2$As ingredient were replaced by Co$_2$As. The mixture of ~1 g was pressed into a cubic pellet, then, sintered at 900°C for 20 h in an evacuated silica tube. Electrical resistivity measurements were conducted at 1.9–300 K by a DC four-probe technique using a Physical Properties Measurement System (Quantum Design Inc.).

High-resolution synchrotron XRD measurements at temperatures from 30 K to 300 K were performed at the BL02B2 beamline in the SPring-8 Japan [18]. Polycrystalline $Ae$Fe$_{1-x}$Co$_x$AsF samples were ground into a fine powder with few tens of micrometers in size, then each of them were packed and sealed in a Lindemann glass capillary with 0.3 mm inner diameter. The capillary was set at a sample stage of a large Debye-Scherrer camera with a 286.48 mm camera radius. The monochromatic X-ray with wavelength of ~0.05 nm was irradiated, and two-dimensional Debye-Scherrer images were detected by Imaging Plates. For low-temperature measurements, the capillaries were cooled by gas-flow methods using N$_2$ gas for above 93 K or He gas for below 93 K. Diffraction patterns were measured at 300, 200, 100 and 30 K. Smaller temperature step measurements were conducted around $T_{\text{anom}}$ for unsubstituted samples to examine their structural phase transitions.

The diffraction data ranging from 2 to 73° (N$_2$ gas cooling) or to 53° (He gas cooling) with a 0.01° step in 2θ, which corresponds to 0.042 nm and 0.056 nm resolution, respectively, were employed for Le-Bail and Rietveld analyses. The lattice constants of $Ae$Fe$_{1-x}$Co$_x$AsF at various temperatures were refined by Le-Bail method to evaluate their temperature dependency. Several weak extrinsic peaks attributed to FeAs, Fe$_2$As and CaF$_2$ (or SrF$_2$) were fitted as impurity phases in each analysis. Further, the diffraction patterns measured at room temperature were subjected to Rietveld analysis to examine the structural differences between CaFeAsF and SrFeAsF and the Co-substituted effects on their structures. In the analyses, the simulated patterns of the CaFeAsF were not fitted sufficiently well to the observed diffraction patterns due presumable to the $c$-axis preferred-orientations, especially in the 12% Co-substituted sample. The fitting results and their $R$ values were substantially improved by using Sasa-Uda function [19] for the preferred-orientation correction. On the other hand, such preferred-orientation was not observed in the SrFe$_{1-x}$Co$_x$AsF samples and their diffraction patterns were

## 3. Results and Discussion

### 3.1. Temperature-dependent resistivity

Figure 1 shows temperature-dependent electrical resistivity of $CaFe_{1-x}Co_xAsF$ (figure 1(a)) and $SrFe_{1-x}Co_xAsF$ (figure 1(b)) for $x$ = 0, 0.06, and 0.12. Unsubstituted samples ($x$ = 0) exhibit large decreases in the ρ-$T$ curves at $T_{anom}$ of ~120 K for CaFeAsF and ~180 K for SrFeAsF. Above $T_{anom}$, the resistivity decreases with $x$ value at a fixed temperature in both compounds, suggesting the conductive careers are enhanced as a result of the substitution. The reduction in the resistivity in CaFeAsF system was smaller than that of SrFeAsF at the same $x$ value. It is noteworthy that $CaFe_{1-x}Co_xAsF$ exhibits superconductivity at $T_c$ ~7 K for $x$ = 0.06 and ~23 K for $x$ = 0.12. On the other hand, the 6% and 12% Co-substituted SrFeAsF samples do not show any clear superconductivity down to 2 K although small decreases in the resistivity were observed in both samples below 10 K. In addition, 6% Co-substituted CaFeAsF exhibits a small shoulder at ~55 K in the ρ-$T$ curve, suggesting the superconductivity appears in the orthorhombic phase. A similar resistivity shoulder is seen at ~70 K in the 6% Co-substituted SrFeAsF although its shape is not clear.

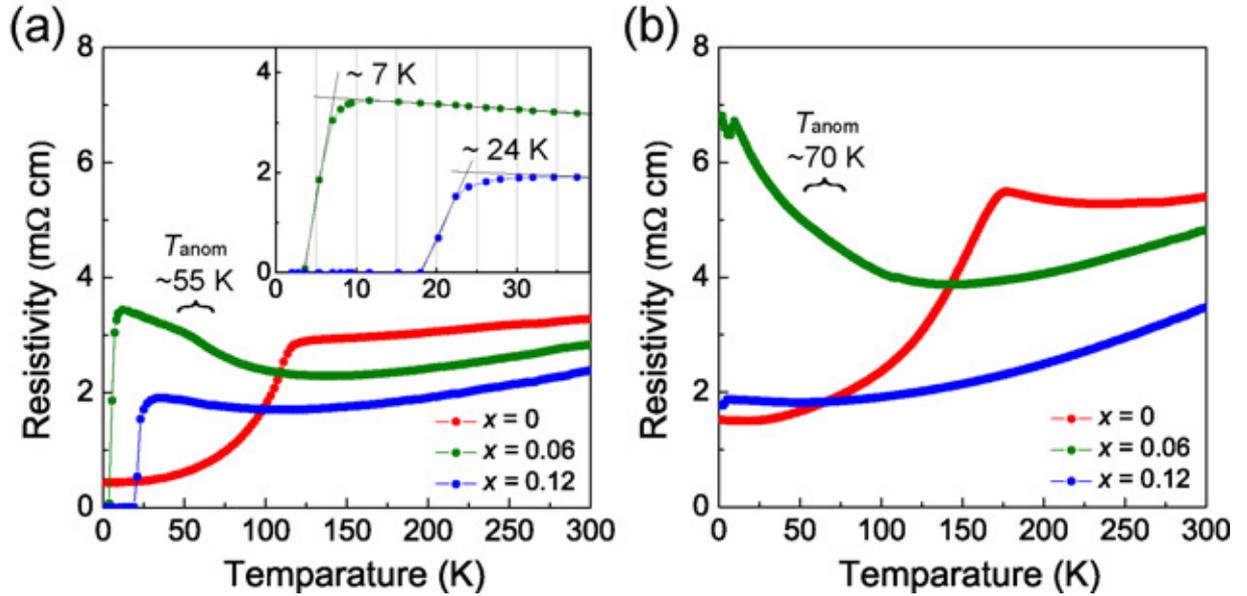

**Figure 1.** Temperature-dependent resistivity of $CaFe_{1-x}Co_xAsF$ (a) and $SrFe_{1-x}Co_xAsF$ (b) ($x$ = 0, 0.06, 0.12). The insert of (a) is a magnified view of 0–40 K.

*3.2. XRD measurements*

*3.2.1. Structural phase transition in unsubstituted and 6% Co-substituted AeFeAsF*

XRD peak profiles of the tetragonal 220 and 214 reflections of unsubstituted samples at temperatures around $T_{anom}$ are shown in figure 2(a) for CaFeAsF and in figure 2(b) for SrFeAsF as typical examples. Both the 220 reflections split into doublet peaks below 120 K for CaFeAsF and 185 K for SrFeAsF, assigned as the orthorhombic 400 and 040 reflections. All the diffraction peaks in the XRD patterns of the unsubstituted CaFeAsF and SrFeAsF samples at the higher temperatures are fitted with the tetragonal *P*4/*nmm* space group with the ZrCuSiAs-type structure, whereas those at the lower temperatures are with the orthorhombic *Cmma* space group, the space groups being the same as those of the unsubstituted LaFeAsO in both temperature regions [9].

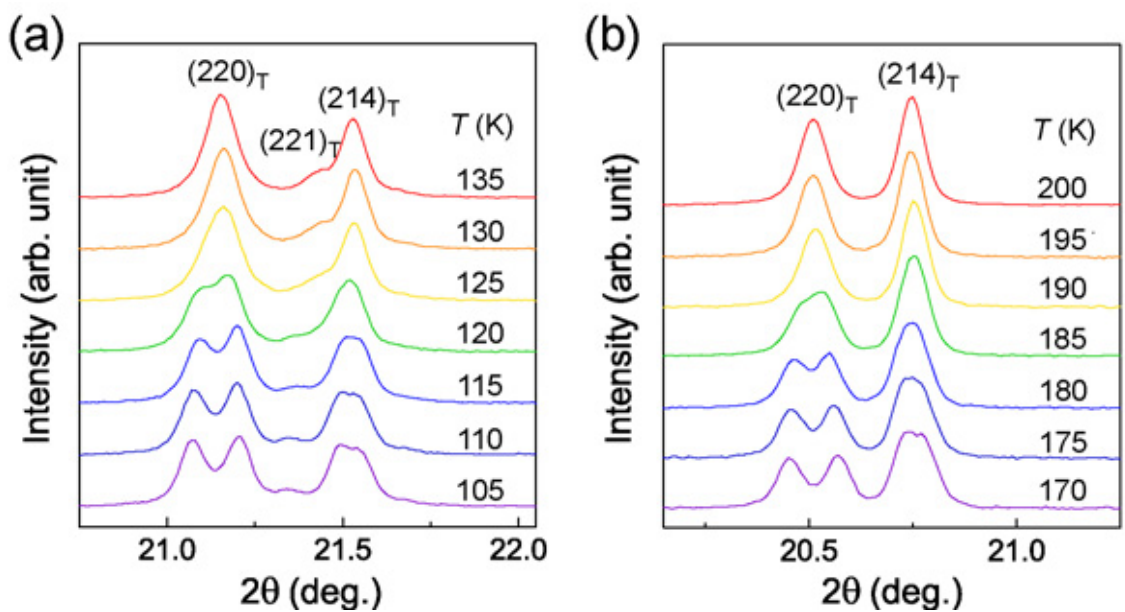

**Figure 2.** Temperature-dependent XRD peak profiles of unsubstituted CaFeAsF (a) and SrFeAsF (b) at around $T_{anom}$. The tetragonal 220 and 214 reflections in both compounds split into the orthorhombic 400, 040 and 134, 314 reflections below $T_{anom}$.

Figure 3 shows the peak profiles of the tetragonal 220 and 214 reflections of Co-substituted CaFeAsF (figures 3(a) and 3(c)) and SrFeAsF (figures 3(b) and 3(d)) at several temperatures. It is noteworthy that the 220 reflection of the 6% Co-substituted CaFeAsF exhibit a single peak down to 65 K, but split into doublet peaks at 30 K (figure 3(a)). The 220 reflection of 6% Co-substituted SrFeAsF also exhibits the remarkable broadening at 30 K (figure 3(b)) although it does not clearly split into two peaks. The Rietveld fitting results of both 6% Co-substituted samples at 30 K are not sufficient by using the *P*4/*nmm* space group but substantially improved by using the *Cmma* space group. On the other hand, both 12% Co-substituted CaFeAsF and SrFeAsF samples (figures 3(c) and 3(d)) show neither splitting nor broadening down to 30 K and the patterns are consistently fitted with the *P*4/*nmm* space group in the whole temperature region examined.

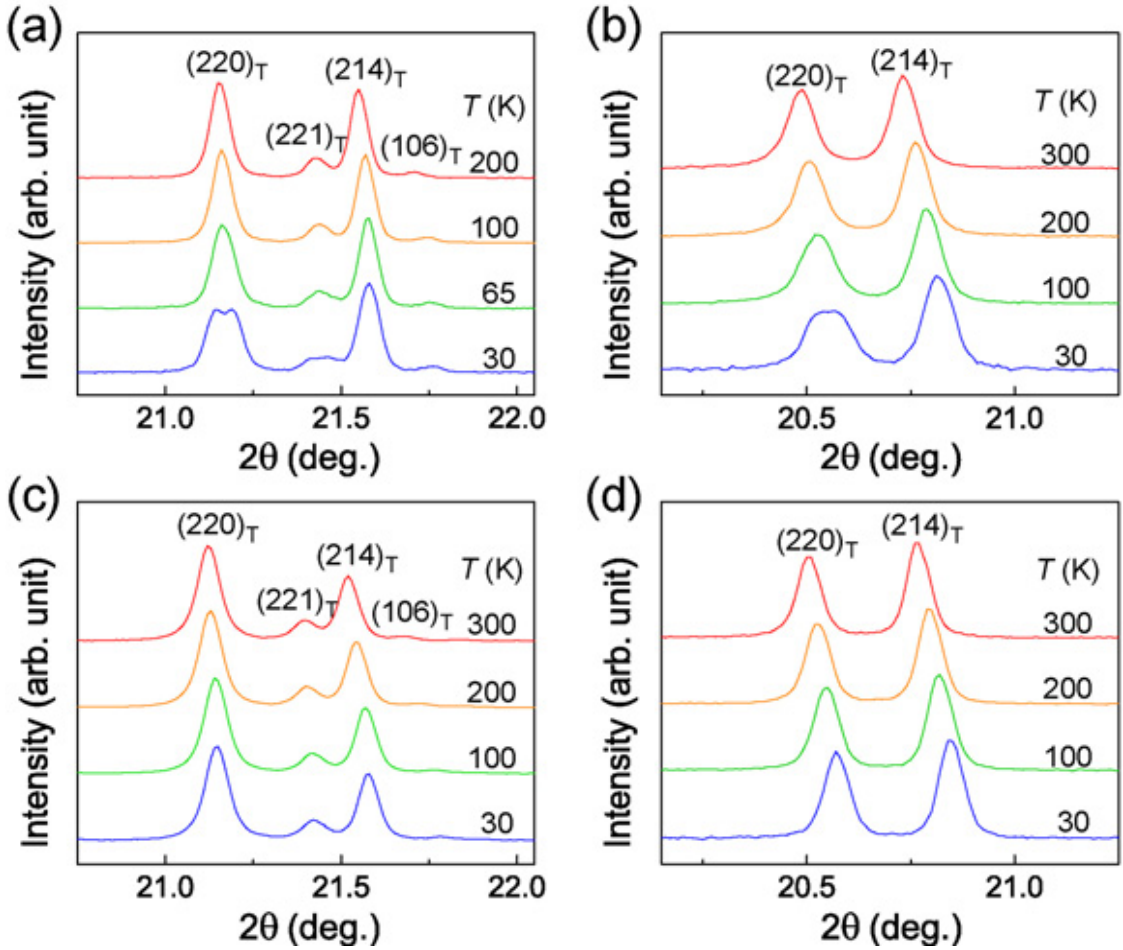

**Figure 3.** Temperature-dependent XRD peak profiles of $CaFe_{1-x}Co_xAsF$ ($x$ = 0.06 (a) and 0.12 (c)) and $SrFe_{1-x}Co_xAsF$ ($x$ = 0.06 (b) and 0.12 (d)) at several temperatures.

The lattice constants of $CaFe_{1-x}Co_xAsF$ and $SrFe_{1-x}Co_xAsF$ as a function of temperature are summarized in figures 4(a) and 4(b), confirming the existence of the phase transition in unsubstituted and 6% Co-substituted samples and the suppression of the transition by the 12% substitution. In the unsubstituted samples, the temperatures where the splitting starts to occur (120 K for CaFeAsF and 185

K for SrFeAsF) agree with $T_{anom}$ observed in their ρ-$T$ curves. Similarly, the temperatures where the splitting occurs in the 6% Co-substituted samples are consistent with the shoulder appearing temperature (~55 K for CaFeAsF and ~70 K for SrFeAsF). These results suggest that the superconductivity in the 6% Co-substituted CaFeAsF occurs in the orthorhombic phase. They also indicate that the 6% Co-substituted SrFeAsF undergoes the structural phase transition at 30 K although the superconductivity is absent.

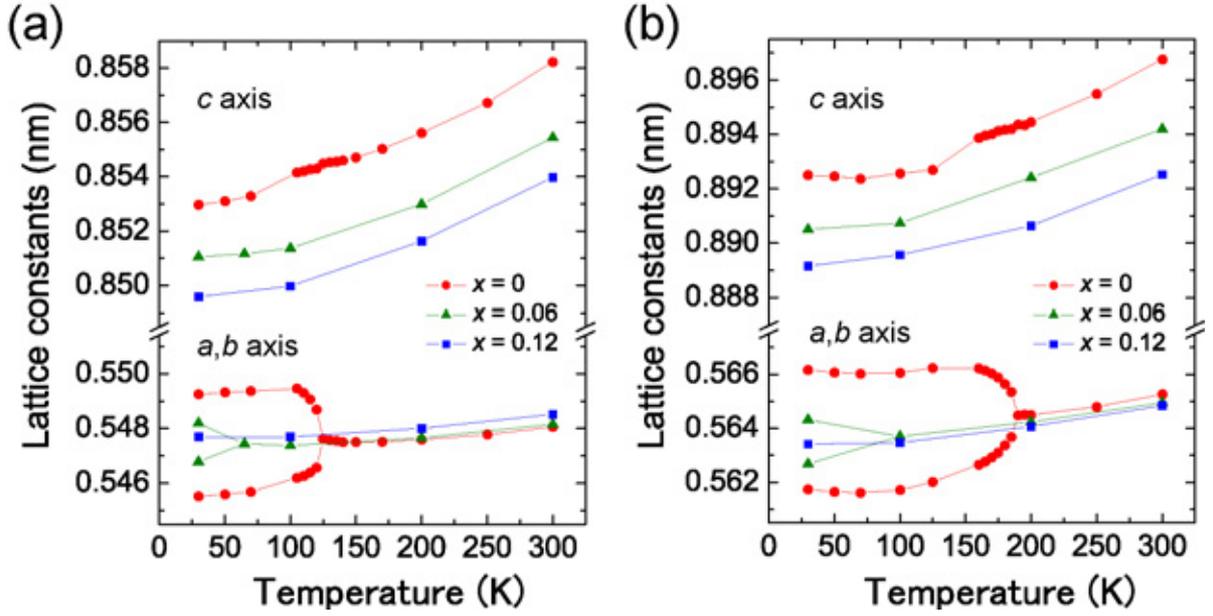

**Figure 4.** Temperature-dependent lattice constants of CaFe$_{1-x}$Co$_x$AsF (a) and SrFe$_{1-x}$Co$_x$AsF (b). The $a$- and $b$-axis lengths in the tetragonal phases above $T_{anom}$ are multiplied by √2.

It is further observed in the figures that the Co substitution largely reduces the $c$-axis length in both CaFeAsF and SrFeAsF, whereas the $a$-axis length slightly changes above $T_{anom}$. For instance, the 12% Co substitution reduces the $c$-axis by ~0.5% and the $a$-axis by ~0.1% in both CaFeAsF and SrFeAsF. The difference between the $a$- and $b$-axes as a result of the phase transition reaches to ~0.7% for CaFeAsF and ~0.8% for SrFeAsF at 30 K, which are much larger than that in the $Ln$FeAsO (~0.5% both for LaFeAsO and SmFeAsO [9,20]). It is also noted that all the lattice constants of unsubstituted CaFeAsF in both the tetragonal and orthorhombic phases monotonically decrease with lowering temperature down to 30 K, whereas those of unsubstituted SrFeAsF decrease down to 70 K, but they increase from 70 K to 30 K in the orthorhombic phase. Such the negative thermal expansion of SrFeAsF has been demonstrated [17] and also reported in unsubstituted PrFeAsO [21], although physical mechanism for it remains unclear.

*3.2.2. Comparison in crystal structure between CaFeAsF and SrFeAsF*

Next, we performed Rietveld analyses on the XRD patterns of CaFe$_{1-x}$Co$_x$AsF and SrFe$_{1-x}$Co$_x$AsF at room temperature to obtain the refined structural parameters. The Rietveld fitting results and the refined structural parameters are shown in online supplementary information. Variations in the bond lengths of Ca–As, Ca–F and Fe–As with Co concentration in the CaFe$_{1-x}$Co$_x$AsF and those of Sr–As, Sr–F and Fe–As in the SrFe$_{1-x}$Co$_x$AsF are shown in figures 5(a) and 5(b). The differences in the bond lengths of Ca–F from Sr–F and Ca–As from Sr–As (~0.013 nm) are in a good agreement with that in the ionic radius between Ca$^{2+}$ ion (0.112 nm) and Sr$^{2+}$ ion (0.125 nm) in the tetrahedral coordination. Further, the Ca-F (~0.234 nm) and Sr-F (~0.247 nm) distances, respectively, are a little smaller than those of CaF$_2$ (~0.237 nm) and SrF$_2$ (~0.251 nm) [22]. The Fe–As distance in unsubstituted CaFeAsF (~0.239 nm) differs from that in unsubstituted SrFeAsF (~0.241 nm) by ~1%. Whereas the Fe–Fe distance, which are equal to the half of the orthorhombic *a*-axis length, is ~0.274 nm for CaFeAsF and ~0.283 nm for SrFeAsF, corresponding to ~3% difference, resulting in the different magnitude of the distortion in the FeAs$_4$ tetrahedron from the regular shape between two compounds as discussed below.

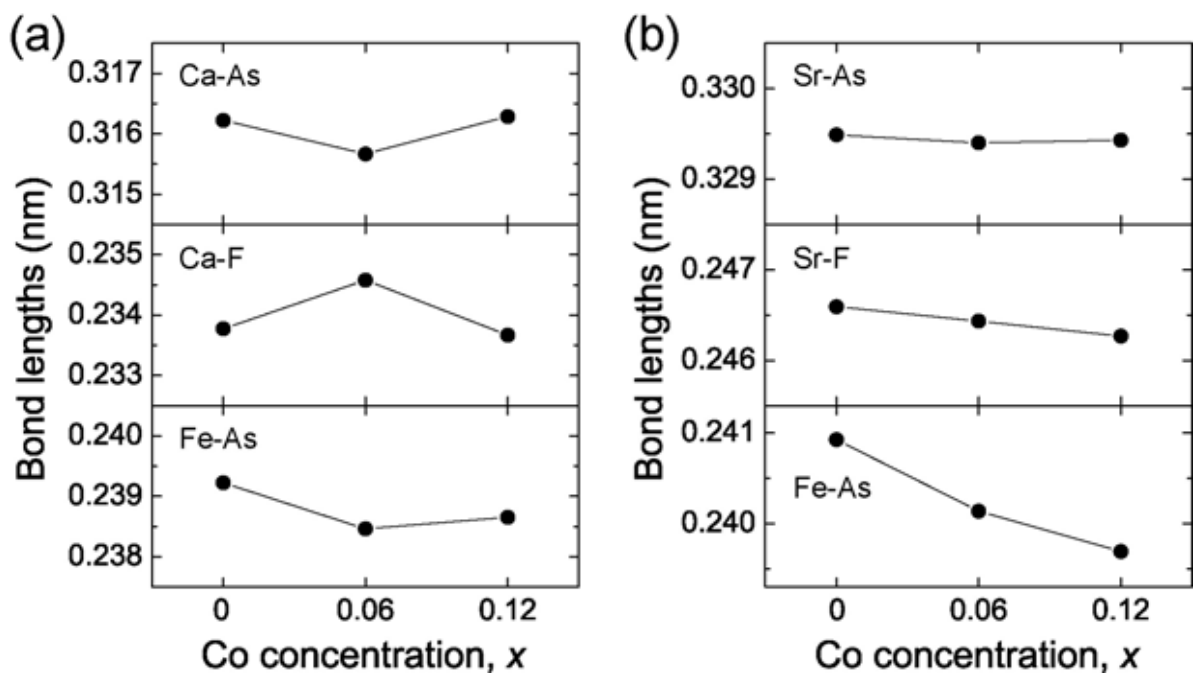

**Figure 5.** Bond lengths of CaFe$_{1-x}$Co$_x$AsF (a) and SrFe$_{1-x}$Co$_x$AsF (b) as a function of Co concentration (*x*). The error bars are comparable to or smaller than the size of plotting symbols.

Figure 6 shows the Co concentration dependence of the As–Fe–As, $Ae$–F–$Ae$ bond angles of FeAs$_4$ and $Ae_4$F tetrahedrons in CaFeAsF and SrFeAsF. The resultant distorted tetrahedrons were schematically illustrated in figure 7. The FeAs$_4$ tetrahedron in unsubstituted CaFeAsF is close to the regular tetrahedron (i.e. 109.47º) with a little elongation along the *c*-axis, while the Ca$_4$F tetrahedron has a large distortion, compressed along the *c*-axis. On the other hand, the FeAs$_4$ tetrahedron in unsubstituted SrFeAsF is significantly compressed, while the Sr$_4$F tetrahedron exhibits a little elongated regular shape along the *c*-axis. That is, the FeAs$_4$ and Ca$_4$F tetrahedrons in CaFeAsF are oppositely distorted against corresponding ones in SrFeAsF, which may be caused by the difference in ionic radii between Ca and Sr. Provided that all the tetrahedrons are regular and the bond lengths are determined only by the ionic radii, the edge length of the Ca$_4$F tetrahedron is ~2.3% shorter than that of FeAs$_4$ tetrahedron in CaFeAsF, whereas that of the Sr$_4$F tetrahedron is ~2.4% longer than that of FeAs$_4$ tetrahedron in SrFeAsF. Therefore, to adjust these mismatches, the Ca$_4$F tetrahedron is flattened by the tension along the *a-b* plane and the FeAs$_4$ tetrahedron is elongated to the *c*-direction by the compression along the *a-b* plane in CaFeAsF. On the other hand, the distortion directions and the shapes of tetrahedrons are opposite in SrFeAsF due to the large Sr$_4$F tetrahedron. Such structural mismatch can be relaxed mainly by modification of the bond angles (i.e. tetrahedron shape) rather than the bond length because of the ionic bond nature in each layer.

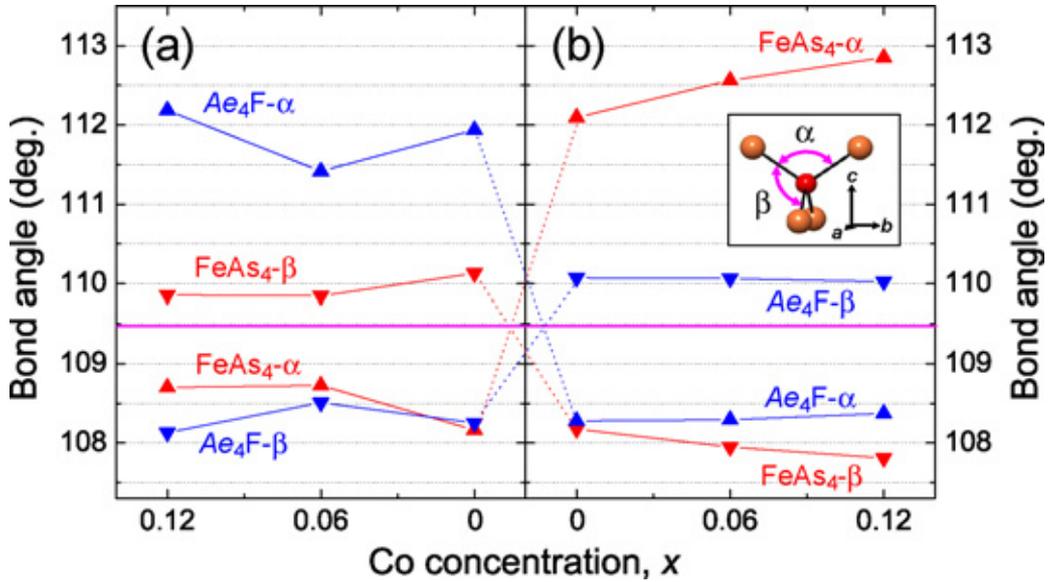

**Figure 6.** Bond angles of the FeAs$_4$ and $Ae_4$F tetrahedrons as a function of Co concentration (*x*) for CaFe$_{1-x}$Co$_x$AsF (a) and SrFe$_{1-x}$Co$_x$AsF (b). Red and blue triangles correspond to As-Fe-As and $Ae$-F-$Ae$ bond angles, and pink line indicates the bond angle of regular tetrahedron (i.e. 109.47°). Inset represents the tetrahedron angles, α and β. The error bars are comparable to or smaller than the size of plotting

symbols. Notice that Co concentration increases toward left in figure 6(a), toward right in figure 6(b) for comparison of each tetrahedron angle of unsubstituted samples.

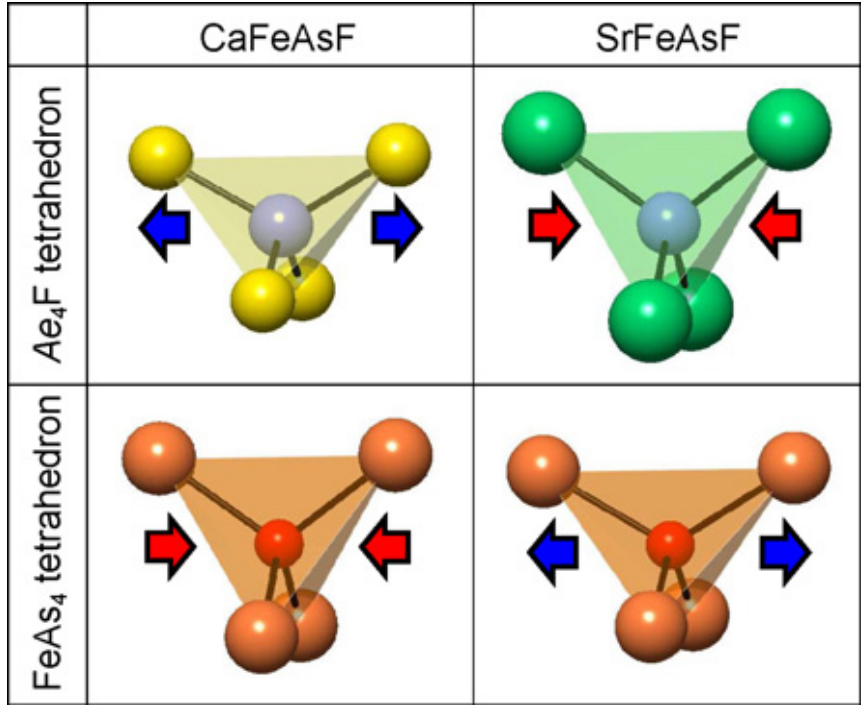

**Figure 7.** Schematic diagram for the $Ae_4$F and FeAs$_4$ tetrahedrons of CaFeAsF and SrFeAsF. The bond angles are emphasized for lucidity. Arrows indicate the directions in which tetrahedrons are distorted due to the different tetrahedron size with that of the other layer.

The geometric discrepancy between the layers is released by the structural relaxation in the tetrahedrons, nevertheless, the different magnitude of the distortion in the FeAs$_4$ tetrahedron results in ~3% difference in Fe–Fe distance between CaFeAsF and SrFeAsF as mentioned above. The Fe–Fe distance in the square lattice is an important parameter which influences on the spin and charge interactions between Fe elements, most likely leading to a change in their $T_c$. For instance, the Fe–Fe distances of LaFeAsO ($T_c$ ~26 K) and SmFeAsO ($T_c$ ~55 K) are ~0.285 nm and ~0.278 nm, which are close to that of SrFeAsF and CaFeAsF, respectively. The difference may result in the reported higher $T_c$ in the Co-substituted CaFeAsF (22 K) than that in the Co-substituted SrFeAsF (7 K) [15,16]. It is also considered that the distortion of the FeAs$_4$ tetrahedron from the regular one is smaller in CaFeAsF than that in SrFeAsF, which may lead to the higher $T_c$ in CaFeAsF than that in SrFeAsF as suggested by Lee et al. [23], that is, the parent compounds of $Ae$FeAsF with the less distorted FeAs$_4$ tetrahedron tends to exhibit higher $T_c$ in the Co-substituted superconductors.

*3.2.3. Co substitution effects on crystal structure of AeFeAsF*

The substitution of Co ion at the Fe site monotonically reduces the Fe–As bond length in the FeAs layer of SrFe$_{1-x}$Co$_x$AsF (figure 5(b)), and the reduction reaches ~0.6% by the 12% Co substitution. Further, it enhances the deviation of the As–Fe–As angle from that of the regular tetrahedron of 109.47º (figure 6(b)). In contrast, the Co substitution slightly modifies the Sr–F bond length (~0.1% decrease by the 12% Co substitution) and the Sr–F–Sr bond angles in the SrF layer. In addition, the interlayer (Sr–As bond length) distance hardly changes by the Co substitution. Such structural changes are much different from those in F-substituted LaFeAsO [9,24], where the F substitution relaxes the distortion of the La$_4$O tetrahedron structure and decreases the distance between the layers (La–As bond length), but it affects slightly on the FeAs$_4$ tetrahedron. The differences may originate, first, from the fact that the substituted layer suffers the structural modification more prominently compared to the other layer. Second, the Co substitution in the FeAs layer increases only the carrier density of the layer, while the indirect doping enhances not only the carrier density, but the charge polarization of the layers as well. It enhances Coulomb interaction between the layers, leading to the larger reduction in the layer distance and the *c*-axis length.

In the case of CaFeAsF, the bond lengths and angles (figures 5(a) and 6(a)) seems to not change systematically with the Co concentration comparing with the case of SrFeAsF. It may be due to the *c*-axis preferred-orientation of CaFe$_{1-x}$Co$_x$AsF powder samples as described in the experimental session. Nevertheless, the analyzed results are still reliable enough for qualitative discussion. It is noteworthy that the Fe–As distances tends to decrease by Co substitution similar to SrFe$_{1-x}$Co$_x$AsF, whereas the substitution changes the As–Fe–As angle toward the regular angle, which is contrary to SrFe$_{1-x}$Co$_x$AsF. The Co substitution induces a modification of the FeAs tetrahedron into more flattened one in both CaFeAsF and SrFeAsF, leading to a change in the originally elongated tetrahedron toward regular one in CaFeAsF and the flattened tetrahedron toward more distorted one in SrFeAsF. Since it is recognized that Fe-based superconductors having the smaller distortion in the FeAs$_4$ tetrahedron tend to exhibit higher $T_c$, the structural analyses suggests that CaFeAsF is a possible candidate for a higher-$T_c$ superconductor provided that the indirect carrier doping techniques are realized. This suggestion is further supported by the fact that CaFeAsF exhibits the highest $T_c$ among the Co-substituted Fe-1111 compounds including LaFe$_{1-x}$Co$_x$AsO and SmFe$_{1-x}$Co$_x$AsO [26,27].

## 4. Summary

We measured powder synchrotron XRD patterns of newly found members of the Fe-1111 superconductor family, CaFe$_{1-x}$Co$_x$AsF and SrFe$_{1-x}$Co$_x$AsF ($x$ = 0, 0.06, 0.12) from 30 K to 300 K. The refined crystal structures were evaluated with the aid of Le-Bail and Rietveld analyses. Obtained results are summarized as follows:

1. The tetragonal to orthorhombic phase transitions were observed at ~120 K for unsubstituted CaFeAsF and at ~180 K for unsubstituted SrFeAsF and the transition temperatures were in a good agreement with kinks observed in the temperature-dependent resistivity curves. The transition temperature decreases with $x$, but the transition sustains with the Co substitution to $x$ = 0.06 in both systems.

2. The layer spacing of SrFeAsF seems to be not altered largely with the Co content, which is different from the case of F-substituted LaFeAsO due presumable to a slight enhancement of charge polarization of the layers by the direct doping.

3. Distortion of the FeAs$_4$ tetrahedron from the regular tetrahedron likely originates from mismatches in atomic radii among the constituent elements. For instance, a larger radius value of the Sr than that of Ca causes a flattened distortion of the FeAs$_4$ tetrahedron in SrFeAsF, which is further enhanced by substitution of the Fe for Co. Whereas, Co substitution to CaFeAsF leads the slightly-elongated FeAs$_4$ tetrahedron to the more regular shape, which may be a reason why the CaFeAsF exhibits the higher $T_c$. We suggest that CaFeAsF is a possible candidate for a higher-$T_c$ superconductor due to its regular shape of FeAs tetrahedron and the highest $T_c$ among the Co-substituted Fe-1111 compounds.

**Acknowledgements**

The authors thank Dr. Y. Kamihara for experimental supports and helpful discussion. We are also grateful to Prof. Y. Kubota of Osaka Prefecture University for support for the XRD analysis.

**Supplementary information for "Comparison of crystal structures and effects of Co substitution in a new member of Fe-1111 superconductor family *Ae*FeAsF (*Ae* = Ca and Sr): a possible candidate for higher-$T_c$ superconductor"**

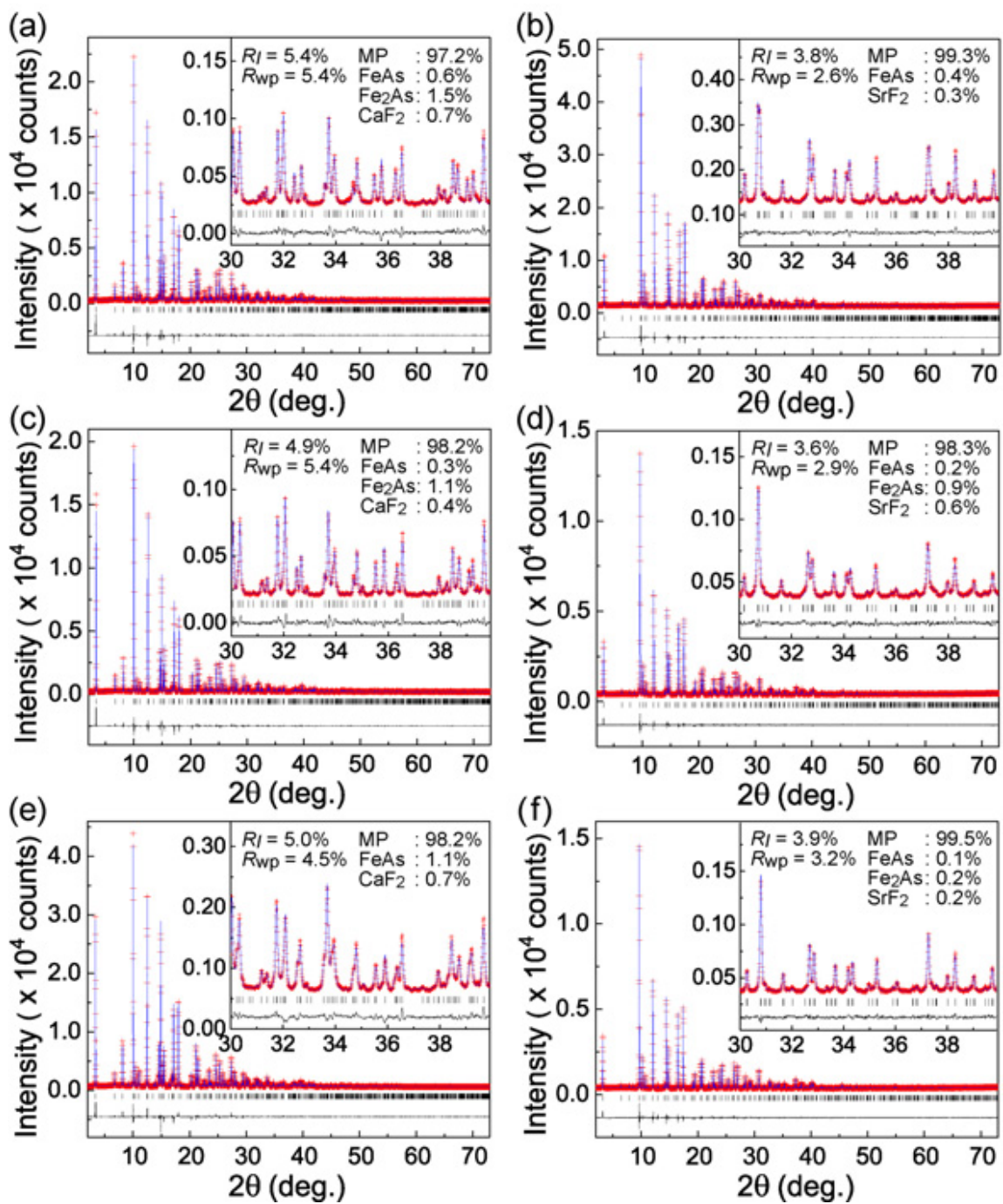

**Figure S1.** Rietveld fitting results of CaFe$_{1-x}$Co$_x$AsF (*x* = 0(a), 0.06(c), 0.12(e)) and SrFe$_{1-x}$Co$_x$AsF (*x* = 0(b), 0.06(d), 0.12(f)) at 300 K. Insets show magnified views from 30 to 40 degrees. R factors and mass percentages of main phase (MP) and impurity phases estimated by the Rietveld analyses are listed. The peak positions of the main phase are indicated below each XRD pattern.

**Table S1.** The structure parameters determined by Rietveld anlaysis for CaFe$_{1-x}$As$_x$F with $x$ = 0(a), 0.06(b), 0.12(c) at 300 K. The occupancy of the Co atom was fixed to each $x$ value and the isotropic atomic displacement parameter ($B$) is constrained to be that of Fe atom. It must be notice that these parameters are tentative because diffraction patterns of CaFe$_{1-x}$As$_x$F more or less suffer influences due to the preferred-orientation as mentioned in the main text.

| Atom | site | Occ. | $x$ | $y$ | $z$ | $B$(Å$^2$) |
|---|---|---|---|---|---|---|
| (a) CaFeAsF at 300 K ($P4/nmm$, $Z$ = 2) | | | | | | |
| $a$ = 0.387471(2) nm, $c$ = 0.858118(6) nm, $V$ = 0.128833(2) nm$^3$ | | | | | | |
| Ca | 2$c$ | 1.000 | 0.25 | 0.25 | 0.15220(15) | 0.82(2) |
| Fe | 2$b$ | 1.000 | 0.75 | 0.25 | 0.5 | 0.503(14) |
| As | 2$c$ | 1.000 | 0.25 | 0.25 | 0.66369(8) | 0.610(11) |
| F | 2$a$ | 1.000 | 0.75 | 0.25 | 0 | 1.22(6) |
| (b) CaFe$_{0.94}$Co$_{0.06}$AsF at 300 K ($P4/nmm$, $Z$ = 2) | | | | | | |
| $a$ = 0.387605(2) nm, $c$ = 0.855441(6) nm, $V$ = 0.128519(2) nm$^3$ | | | | | | |
| Ca | 2$c$ | 1.000 | 0.25 | 0.25 | 0.15413(14) | 0.86(2) |
| Fe | 2$b$ | 0.940 | 0.75 | 0.25 | 0.5 | 0.593(14) |
| Co | 2$b$ | 0.060 | 0.75 | 0.25 | 0.5 | 0.593(14) |
| As | 2$c$ | 1.000 | 0.25 | 0.25 | 0.66243(7) | 0.657(10) |
| F | 2$a$ | 1.000 | 0.75 | 0.25 | 0 | 1.20(7) |
| (c) CaFe$_{0.88}$Co$_{0.12}$AsF at 300 K ($P4/nmm$, $Z$ = 2) | | | | | | |
| $a$ = 0.387861(2) Å, $c$ = 0.853974(7) Å, $V$ = 0.128469(2) Å$^3$ | | | | | | |
| Ca | 2$c$ | 1.000 | 0.25 | 0.25 | 0.15264(14) | 0.73(2) |
| Fe | 2$b$ | 0.880 | 0.75 | 0.25 | 0.5 | 0.520(14) |
| Co | 2$b$ | 0.120 | 0.75 | 0.25 | 0.5 | 0.520(14) |
| As | 2$c$ | 1.000 | 0.25 | 0.25 | 0.66288(7) | 0.576(10) |
| F | 2$a$ | 1.000 | 0.75 | 0.25 | 0 | 1.42(7) |

**Table S2.** The structure parameters determined by Rietveld anlaysis for SrFe$_{1-x}$As$_x$F with $x = 0$(a), 0.06(b), 0.12(c) at 300 K. The occupancy of the Co atom was fixed to each $x$ value and the isotropic atomic displacement parameter ($B$) is constrained to be that of Fe atom.

| Atom | site | Occ. | $x$ | $y$ | $z$ | $B(Å^2)$ |
|---|---|---|---|---|---|---|
| (a) SrFeAsF at 300 K ($P4/nmm, Z = 2$) | | | | | | |
| $a = 0.399697(2)$ nm, $c = 0.896769(7)$ nm, $V = 0.143266(3)$ nm$^3$ | | | | | | |
| Sr | 2c | 1.000 | 0.25 | 0.25 | 0.16109(8) | 0.85(2) |
| Fe | 2b | 1.000 | 0.75 | 0.25 | 0.5 | 0.74(2) |
| As | 2c | 1.000 | 0.25 | 0.25 | 0.65005(10) | 0.50(2) |
| F | 2a | 1.000 | 0.75 | 0.25 | 0 | 1.15(9) |
| (b) SrFe$_{0.94}$Co$_{0.06}$AsF at 300 K ($P4/nmm, Z = 2$) | | | | | | |
| $a = 0.399477(2)$ nm, $c = 0.894197(8)$ nm, $V = 0.142698(3)$ nm$^3$ | | | | | | |
| Sr | 2c | 1.000 | 0.25 | 0.25 | 0.16142(9) | 0.84(2) |
| Fe | 2b | 0.940 | 0.75 | 0.25 | 0.5 | 0.62(3) |
| Co | 2b | 0.060 | 0.75 | 0.25 | 0.5 | 0.62(3) |
| As | 2c | 1.000 | 0.25 | 0.25 | 0.64908(10) | 0.44(2) |
| F | 2a | 1.000 | 0.75 | 0.25 | 0 | 1.00(9) |
| (c) SrFe$_{0.88}$Co$_{0.12}$AsF at 300 K ($P4/nmm, Z = 2$) | | | | | | |
| $a = 0.399427(2)$ nm, $c = 0.892577(7)$ nm, $V = 0.142403(3)$ nm$^3$ | | | | | | |
| Sr | 2c | 1.000 | 0.25 | 0.25 | 0.16146(4) | 0.75(2) |
| Fe | 2b | 0.880 | 0.75 | 0.25 | 0.5 | 0.68(3) |
| Co | 2b | 0.120 | 0.75 | 0.25 | 0.5 | 0.68(3) |
| As | 2c | 1.000 | 0.25 | 0.25 | 0.64851(10) | 0.53(2) |
| F | 2a | 1.000 | 0.75 | 0.25 | 0 | 1.10(10) |